\documentclass[12pt]{article}
\usepackage{a4wide,epsfig,psfrag,amsmath,amssymb,cite,scalefnt}
\usepackage{color}
\usepackage{rotating}

\graphicspath{ {figs/} }

\newcommand{\qsla}{q\!\!\!/\,\,}
\newcommand{\Qsla}{Q\!\!\!\!/\,\,\,}
\newcommand{\order}[1]{\mathcal{O}\left( #1 \right)}

\newcommand{\cR}{C_F}
\newcommand{\cA}{C_A}
\newcommand{\tr}{T}
\newcommand{\xinl}{\xi^{(n_l)}}

\allowdisplaybreaks

\begin{document}

\title{\vskip-3cm{\baselineskip14pt
    \begin{flushleft}
      \normalsize DESY 18-011, TTP18-007
    \end{flushleft}} \vskip1.5cm 
  Four-loop wave function renormalization\\ in QCD and QED
}

\author{
  Peter Marquard$^{a}$,
  Alexander V. Smirnov$^{b}$,
  \\
  Vladimir A. Smirnov$^{c}$,
  Matthias Steinhauser$^{d}$
  \\[1em]
  {\small\it (a) Deutsches Elektronen-Synchrotron, DESY}
  {\small\it  15738 Zeuthen, Germany}
  \\
  {\small\it (b) Research Computing Center, Moscow State University}\\
  {\small\it 119991, Moscow, Russia}
  \\  
  {\small\it (c) Skobeltsyn Institute of Nuclear Physics of Moscow State University}\\
  {\small\it 119991, Moscow, Russia}
  \\
  {\small\it (d) Institut f{\"u}r Theoretische Teilchenphysik,
    Karlsruhe Institute of Technology (KIT)}\\
  {\small\it 76128 Karlsruhe, Germany}  
}
  
\date{}

\maketitle

\thispagestyle{empty}

\begin{abstract}

  We compute the on-shell wave function renormalization constant to four-loop order in
  QCD and present numerical results for all coefficients of the SU$(N_c)$
  colour factors.  We extract the four-loop HQET anomalous dimension of the
  heavy quark field and also discuss the application of our result to QED.

%

\end{abstract}

\thispagestyle{empty}


\newpage


\section{Introduction}

Heavy quarks play an important role in modern particle physics,
in particular in the context of Quantum Chromodynamics (QCD).
This concerns both virtual effects, the production of massive quarks
at collider experiments and the study of bound state effects
of heavy quark-anti-quark pairs.

Processes which involve heavy quarks require the renormalization constants for
the heavy quark mass and, when they appear as external particles, also for
the quark wave function. The mass renormalization constant in the on-shell
scheme, $Z_m^{\rm OS}$, has been computed to four-loop order in
Refs.~\cite{Marquard:2015qpa,Marquard:2016dcn}.  In this work we compute the
wave function renormalization constant in the on-shell scheme, $Z_2^{\rm OS}$,
to the same order in perturbation theory. $Z_2^{\rm OS}$
is needed for all processes involving
external heavy quarks to obtain properly normalized Green's functions as
dictated by the Lehmann-Symanzik-Zimmermann (LSZ) reduction formula.
Currently there is no immediate application for the four-loop term of
$Z_2^{\rm OS}$. However, it is an important building block for future
applications. For example, it enters all processes which involve the massive
four-loop form factor. $Z_2^{\rm OS}$ is also needed for the 
five-loop corrections to static properties like the
anomalous magnetic moment of quarks or, in the case of QED, of leptons.

The calculation of $Z_2^{\rm OS}$ is for several reasons more involved
than the one of $Z_m^{\rm OS}$. First of all, one has to compute the
derivative of the fermion self energy which leads to  higher
powers of propagators and thus to a more involved reduction problem.
Furthermore, $Z_2^{\rm OS}$ contains both ultraviolet and infrared
divergences. Thus, dividing $Z_2^{\rm OS}$ by its $\overline{\rm MS}$
counterpart does not lead to a finite quantity as in the case of
$Z_m^{\rm OS}$. $Z_2^{\rm OS}$ also depends on the QCD gauge
parameter whereas $Z_m^{\rm OS}$ does not.

The on-shell renormalization constants $Z_2^{\rm OS}$ and $Z_m^{\rm OS}$
can be extracted from the quark propagator by demanding
that the quark two-point function has a zero at the position of the
on-shell mass and that the residue of the propagator is $-i$.
In the following we briefly sketch the derivation of the relations between
the heavy quark self energy and $Z_2^{\rm OS}$ and $Z_m^{\rm OS}$.

The renormalized quark propagator is given by
\begin{eqnarray}
  S_F(q) &=& \frac{-i Z_2^{\rm OS}}{\qsla - m^0 + \Sigma(q,M)}
  \,,
  \label{eq::defs::fullprop2}
\end{eqnarray}
where the renormalization constants are defined as
\begin{eqnarray}
  m^0 &=& Z_m^{\rm OS}\, M\,,
  \nonumber\\
  \psi^0 &=& \sqrt{Z_2^{\rm OS}}\, \psi\,.
  \label{eq::defs::defZ2}
\end{eqnarray}
$\psi$ is the quark field with mass $m$, $M$ is the 
on-shell mass and bare quantities are denoted by a superscript 0.
$\Sigma$ denotes the quark self energy which is conveniently decomposed as
\begin{eqnarray}
  \Sigma(q,m) &=&
  m\, \Sigma_1(q^2,m) + (\qsla - m)\, \Sigma_2(q^2,m)\,.
  \label{eq::defs::sigmadecomp}
\end{eqnarray}
In the limit $q^2\to M^2$ we require
\begin{eqnarray}
  S_F(q)
  &\stackrel{q^2\to M^2}{\longrightarrow}& \frac{-i}{\qsla - M}\,.
  \label{eq::defs::fullprop2_2}
\end{eqnarray}

The calculation outlined in Ref.~\cite{Gray:1990yh} for the evaluation of
$Z_m^{\rm OS}$ and $Z_2^{\rm OS}$ reduces all occurring Feynman diagrams
to the evaluation of on-shell integrals at the bare mass scale. In
particular, it avoids the introduction of explicit counterterm diagrams.
We find it more convenient to follow the more
direct approach described in
Refs.~\cite{Melnikov:2000qh,Melnikov:2000zc}, which requires 
the calculation of diagrams with mass counterterm insertion. 

Following Refs.~\cite{Gray:1990yh,Melnikov:2000qh,Melnikov:2000zc,Marquard:2007uj}
we expand $\Sigma$ around $q^2 = M^2$ and obtain
\begin{eqnarray}
  \Sigma(q,M) &\approx& M\, \Sigma_1(M^2,M) + (\qsla - M)\,
  \Sigma_2(M^2,M) \nonumber\\
  && + M \frac{\partial}{\partial q^2} \Sigma_1(q^2,M) \Big|_{q^2 =
  M^2} (q^2 - M^2) + \dots \nonumber\\
  &\approx& M\, \Sigma_1(M^2,M) \nonumber\\ 
  && + (\qsla - M) \left( 2M^2 \frac{\partial}{\partial q^2}
  \Sigma_1(q^2,M) \Big|_{q^2 = M^2} + \Sigma_2(M^2,M) \right) +
  \dots\,.
  \label{eq::defs::taylor}
\end{eqnarray}
Inserting Eq.~(\ref{eq::defs::taylor}) into Eq.~(\ref{eq::defs::fullprop2})
and comparing to Eq.~(\ref{eq::defs::fullprop2_2}) leads to the
following formulae for the renormalization constants
\begin{eqnarray}
  Z_m^{\rm OS} &=& 1 + \Sigma_1(M^2,M)\,,
  \nonumber \\
  \left( Z_2^{\rm OS} \right)^{-1} &=& 1 + 2M^2
  \frac{\partial}{\partial q^2} \Sigma_1(q^2,M) \Big|_{q^2 = M^2} +
  \Sigma_2(M^2,M) \,.
  \label{eq::defs::calcZ2}
\end{eqnarray}
Thus, $Z_m^{\rm OS}$ is obtained from $\Sigma_1$ for $q^2 = M^2$.  To
calculate $Z_2^{\rm OS}$, one has to compute the first derivative of the
self-energy diagrams. The mass renormalization is taken into account
iteratively by calculating lower-loop diagrams with zero-momentum
insertions.

It is convenient to introduce $q = Q(1+t)$
with $Q^2 = M^2$ and re-write the self energy as
\begin{equation}
  \Sigma(q,M) = M \Sigma_1(q^2,M) + (\Qsla - M)
  \Sigma_2(q^2,M) + t\Qsla \Sigma_2(q^2,M)\,.
  \label{eq::defs::seldecompt}
\end{equation}
Let us now consider the quantity ${\rm Tr} \{ \frac{\Qsla + M}{4M^2}
\Sigma \}$ and expand it to first order in $t$ which leads to
\begin{eqnarray}
  {\rm Tr} \left\{ \frac{\Qsla + M}{4M^2} \Sigma(q,M) \right\} &=&
  \Sigma_1(q^2,M) + t \Sigma_2(q^2,M) \nonumber \\
  &=& \Sigma_1(M^2,M)
  + \left( 2M^2 \frac{\partial}{\partial q^2} \Sigma_1(q^2,M)
  \Big|_{q^2 = M^2} \!\!+\! \Sigma_2(M^2,M) \right) t \nonumber\\ 
  &&  + \order{t^2} \,.
  \label{eq::defs::trace}
\end{eqnarray}
The comparison to Eq.~(\ref{eq::defs::calcZ2}) shows that the leading term
provides $Z_m^{\rm OS}$ and the coefficient of the linear term in $t$ leads
to $Z_2^{\rm OS}$.

In the next Section we present results for $Z_2^{\rm OS}$ up to four loops and
in Section~\ref{sec::HQET} we discuss consistency checks which
are obtained from matching full QCD to Heavy Quark Effective Theory (HQET).
Section~\ref{sec::concl} contains a brief summary and our conclusions.


\section{\label{sec::Z2OS}Results for $Z_2^{\rm OS}$}

The wave function renormalization constant is conveniently cast into the
form
\begin{eqnarray}
  Z_2^{\rm OS} &=& 1 + 
  \sum_{j\ge1} 
  \left(\frac{\alpha_s^0(\mu)}{\pi}\right)^j 
  \left(\frac{e^{\gamma_E}}{4 \pi} \right)^{-j\epsilon} 
  \left(\frac{\mu^2}{M^2}\right)^{j\epsilon}
  \delta Z_2^{(j)}
  \,,
  \label{eq::deltaZ2}
\end{eqnarray}
where the bare strong coupling constant $\alpha_s^0$ has been used for the
parametrization. 
Note that $\delta Z_2^{(i)}$ for $i\ge3$ depend on the bare QCD gauge parameter
$\xi$ which is introduced in the gluon propagator via
\begin{eqnarray}
  D_g^{\mu\nu}(q) &=& -i\,
  \frac{g^{\mu\nu}- \xi \frac{q^\mu q^\nu}{q^2}}
  {q^2+ i\varepsilon}
  \,.
\end{eqnarray}
With these choices we can define the coefficients $\delta Z_2^{(i)}$ such that
they do not contain $\log(\mu^2/M^2)$ terms. In fact, they can be
combined to the factors $(\mu^2/M^2)^{j\epsilon}$ where $j$ is the loop order
(cf. Eq.~(\ref{eq::deltaZ2})).  The renormalization of $\alpha_s$ and $(\xi-1)$ is
multiplicative so that, if required, $\alpha_s^0$ and $\xi^0$ can be replaced
in a straightforward way by their renormalized counterparts using the
relations
\begin{eqnarray}
  \alpha_s^0 &=& (\mu^2)^{2\epsilon} Z_{\alpha_s} \alpha_s\,,
  \nonumber\\
  \xi^0-1 &=& Z_3 (\xi-1)
  \,,
\end{eqnarray}
where 
\begin{eqnarray}
  Z_{\alpha_s}&=&1+\frac{1}{\epsilon}
  \left(\frac{n_f}{6}-\frac{11}{12}C_A\right)
  \frac{\alpha_s}{\pi}+\ldots\,,\nonumber\\
  Z_3&=&1+\frac{1}{\epsilon}
  \left[-\frac{n_f}{6}+\left(\frac{5}{12}+\frac{1}{8}\xi\right)C_A\right]
  \frac{\alpha_s}{\pi}+\ldots\,.
\end{eqnarray}
$C_A=3$ is a SU(3) colour factor and $n_f$ is the number of active quarks.
The ellipses denote higher order terms in $\alpha_s$.  To obtain the
ultraviolet-renormalized version of $Z_2^{\rm OS}$ we need $Z_{\alpha_s}$ to
three loops and $Z_3$ to one-loop order.  Note that in Eq.~(\ref{eq::deltaZ2})
it is assumed that the heavy quark mass is renormalized on-shell, i.e., all
mass renormalization counterterms from lower-order diagrams are included.

For the calculation of the four-loop diagrams we proceeded in the same way as
for the calculation of the mass renormalization
constant~\cite{Marquard:2015qpa,Marquard:2016dcn} and the muon anomalous 
magnetic moment~\cite{Marquard:2017iib} and thus refer
to~\cite{Marquard:2016dcn} for more details. Let us still describe some
complications.  After a tensor reduction we obtain Feynman integrals from the
same hundred families with 14 indices as
in~\cite{Marquard:2015qpa,Marquard:2016dcn}. The maximal number of positive
indices is eleven. One can describe the complexity of integrals of a given
sector (determined by a decomposition of the set of indices into subsets of
positive and non-positive indices) by the number $\sum |a_i-n_i|$, where the
indices $n_i=1$ or 0 characterize a given sector. What is most crucial for the
feasibility of an integration-by-parts (IBP) reduction is the complexity
of input integrals in the top sector, i.e.  with $n_i=1$ for $i=1,2,\ldots,11$
and $n_i=0$ for $i=12,13,14$.  In the present calculation, this number was up
to six while in our previous calculation it was five. Therefore, the reduction
procedure performed with {\tt
  FIRE}~\cite{Smirnov:2008iw,Smirnov:2013dia,Smirnov:2014hma} coupled with
{\tt LiteRed} \cite{Lee:2012cn,Lee:2013mka} and {\tt Crusher}~\cite{crusher}
was essentially more complicated as compared to that
of~\cite{Marquard:2015qpa,Marquard:2016dcn}.

As in~\cite{Marquard:2015qpa,Marquard:2016dcn} we revealed additional
relations between master integrals of different families using symmetries and
applied the code {\tt tsort} which is part of the latest {\tt FIRE}
version~\cite{Smirnov:2014hma}.  In most cases, the master integrals were
computed numerically with the help of {\tt
  FIESTA}~\cite{Smirnov:2008py,Smirnov:2009pb,Smirnov:2013eza}.  For some
master integrals, we used analytic results obtained by a straightforward
loop-by-loop integration at general dimension $d$ and also used analytical
results obtained for the 13 non-trivial four-loop on-shell master integrals
computed in~\cite{Lee:2013sx}.  As it is described in detail
in~\cite{Marquard:2016dcn}, we also applied Mellin-Barnes
representations~\cite{Smirnov:1999gc,Tausk:1999vh,Czakon:2005rk,Smirnov:2009up}.
In the case of one-fold Mellin-Barnes representations, it is possible to
obtain a very high precision (up to 1000 digits) so that analytic results can
be recovered using the {\tt PSLQ} algorithm~\cite{PSLQ} .  Often the two-,
three- and higher-fold MB representation provide a better precision than {\tt
  FIESTA}. Recently a subset of the master integrals has been calculated either
analytically or with high numerical precision, in the context of the anomalous
magnetic moment of the electron~\cite{Laporta:2017okg}. However, these results
are not available to us.

The more complicated IBP reduction resulted in higher $\epsilon$ poles
in the coefficients of some of the master integrals, so that the corresponding
results are needed to higher
powers in $\epsilon$. Depending on the integral, we either
straightforwardly evaluated more terms with {\tt FIESTA}, or obtained
more analytical terms, or more numerical terms via
Mellin-Barnes integrals.

Let us mention that we compute the self energies on the right-hand
side of Eq.~(\ref{eq::defs::calcZ2}) including terms of order
$\xi^2$. We did not evaluate the $\xi^3$, $\xi^4$ and $\xi^5$
contributions. Some diagrams develop $\xi^6$ terms which we reduced to
master integrals and we could show that their contributions to
$Z_2^{\rm OS}$  add up to zero. Thus, in our final result for
$Z_2^{\rm OS}$ contains $\xi^2$ terms. We cannot exclude that also
higher order $\xi$ terms are present but we do not expect that there
are $\xi^n$ terms present in $Z_2^{\rm OS}$ for $n\ge4$.

Let us in a first step turn to the one-, two- and three-loop results for
 $Z_2^{\rm OS}$ which are available from
Refs.~\cite{Broadhurst:1991fy,Melnikov:2000zc,Marquard:2007uj}. We have added
higher order $\epsilon$ terms which are necessary to obtain $Z_2^{\rm OS}$  at
four loops. In Appendix~\ref{app::Z2OS} we present results which in particular
include the ${\cal O}(\epsilon)$ terms of the three-loop coefficient.

In the following we present results for all 23 SU$(N_c)$ colour
structures which occur at four-loop order. It is convenient to decompose
$\delta Z_2^{(4)}$ as
\begin{eqnarray}
  \delta Z_2^{(4)} &=&
  C_F^4 \,\delta Z_2^{FFFF}
          +  C_F^3 C_A \,\delta Z_2^{FFFA}
          +  C_F^2 C_A^2 \,\delta Z_2^{FFAA}
          +  C_F C_A^3 \,\delta Z_2^{FAAA}
\nonumber\\&&\mbox{}
          +  \frac{d_F^{abcd}d_A^{abcd}}{N_c} \,\delta Z_2^{d_{FA}}
          +  n_l \frac{d_F^{abcd}d_F^{abcd}}{N_c} \,\delta Z_2^{d_{FF}L}
          +  n_h \frac{d_F^{abcd}d_F^{abcd}}{N_c} \,\delta Z_2^{d_{FF}H}
\nonumber\\&&\mbox{}
          +  C_F^3 T n_l \,\delta Z_2^{FFFL}
          +  C_F^2 C_A T n_l \,\delta Z_2^{FFAL}
          +  C_F C_A^2 T n_l \,\delta Z_2^{FAAL}
\nonumber\\&&\mbox{}
          +  C_F^2 T^2 n_l^2 \,\delta Z_2^{FFLL}
          +  C_F C_A T^2 n_l^2 \,\delta Z_2^{FALL}
          +  C_F T^3 n_l^3 \,\delta Z_2^{FLLL}
\nonumber\\&&\mbox{}
          +  C_F^3 T n_h \,\delta Z_2^{FFFH}
          +  C_F^2 C_A T n_h \,\delta Z_2^{FFAH}
          +  C_F C_A^2 T n_h \,\delta Z_2^{FAAH}
\nonumber\\&&\mbox{}
          +  C_F^2 T^2 n_h^2 \,\delta Z_2^{FFHH}
          +  C_F C_A T^2 n_h^2 \,\delta Z_2^{FAHH}
          +  C_F T^3 n_h^3 \,\delta Z_2^{FHHH}
\nonumber\\&&\mbox{}
          +  C_F^2 T^2 n_l n_h \,\delta Z_2^{FFLH}
          +  C_F C_A T^2 n_l n_h \,\delta Z_2^{FALH}
          +  C_F T^3 n_l^2 n_h \,\delta Z_2^{FLLH}
\nonumber\\&&\mbox{}
          +  C_F T^3 n_l n_h^2 \,\delta Z_2^{FLHH}
          \,,
          \label{eq::z24_col}
\end{eqnarray}
where $C_F, C_A, T, n_l$ and $n_h$ are defined after
Eq.~(\ref{eq::deltaZ2OS_3}) in Appendix~\ref{app::Z2OS}.  The new colour
factors at four loops are the symmetrized traces of four generators in the
fundamental and adjoint representation denoted by $d_F^{abcd}$ and
$d_A^{abcd}$, respectively.

In Tabs.~\ref{tab::Z2OSxi0},~\ref{tab::Z2OSxi1},~\ref{tab::Z2OSxi2}
and~\ref{tab::Z2OSxi0noCT} (see Appendix~\ref{app::tables}) we show the
numerical results for the coefficients introduced in Eq.~(\ref{eq::z24_col}).
The numerical uncertainties have been obtained by adding the uncertainties
from each individual master integral in quadrature and multiplying the result
by a security factor~10.  This approach is quite conservative, however, we
observed that there are rare cases where the uncertainty from numerical
integration is underestimated by several standard deviations.  A factor 10
covers all cases which we have experienced (see also discussion in
Ref.~\cite{Marquard:2016dcn}).  All coefficients which have a non-zero
numerical uncertainty are truncated in such a way that two digits of the
uncertainty are shown; otherwise we present (at least) five significant
digits.  Note that the $n_l^3$ and $n_l^2$ terms are known
analytically~\cite{Lee:2013sx}.  None of the other coefficients are known
analytically to us although for some of them the uncertainty is very small,
see, e.g., $C_F n_h^3$.

Let us start with the discussion of Tab.~\ref{tab::Z2OSxi0}.  Most of the
coefficients are known with an uncertainty of a few per cent or below.  An
exception are the $C_F^4$ and $C_F^3 C_A$ colour factors, where the
uncertainty is about 30\%. In the case of $n_h (d_F^{abcd})^2$ the uncertainty
is larger than the central value and we are not able to decide whether
the corresponding coefficient is zero or numerically small.  For some colour
structures our precision is below a per mille level, in particular for the
most non-abelian colour factor $C_F C_A^3$ which provides the numerically
largest contribution.

There are some coefficients in the pole parts where the numerical uncertainty
is larger than the central value.  In these cases no definite conclusion can
be drawn. Within our (conservative) uncertainty estimate the results are
compatible with zero. Still, in these cases we cannot exclude a small non-zero
result.  Note, however, that in most cases the uncertainty is much smaller
than the central value. In particular, all colour structures except those
involving $d_F^{abcd}$ or $d_A^{abcd}$ have a non-zero $1/\epsilon^4$ pole.
In fact, we expect that the colour structures involving
$d_F^{abcd}$ and $d_A^{abcd}$ only have a $1/\epsilon$
poles which is consistent with our result.

The coefficients in Tab.~\ref{tab::Z2OSxi1} representing the linear $\xi$
terms are in general much smaller than for $\xi=0$ and the situation is
similar as for the pole terms of Tab.~\ref{tab::Z2OSxi0}: We can conclude that
the colour structures $C_F^2C_A^2$, $C_FC_A^3$, $d_F^{abcd}d_A^{abcd}$, $C_F
C_A^2 n_l$, $C_F^2 C_A n_h$, $C_F C_A^2 n_h$, $C_F C_A n_h^2$ and $C_F C_A n_l
n_h$ have non-zero coefficients.  Within our precision the coefficient of
$C_F^3C_A$ is zero: the central value is of order $10^{-4}$ and furthermore
ten times smaller than the uncertainty. However, a closer look into this
contribution shows that non-trivial master integrals are involved which
combine to the numerical result given in Tab.~\ref{tab::Z2OSxi1}.  Since the
master integrals
are linear independent and since they are beyond ``three-loop complexity''
(i.e., they are neither products of lower-loop integrals nor contain simple
one-loop insertions) we would expect a non-zero coefficient unless there are
accidental cancellations.
Note that at three-loop order there are two colour structures which have $\xi$
dependent coefficients: $C_F C_A^2$ and $C_F C_A n_h$.

In Tab.~\ref{tab::Z2OSxi2}, which contains the $\xi^2$ terms,
there are non-zero coefficients for the
colour structures $C_FC_A^3$, $d_F^{abcd}d_A^{abcd}$ and $C_F C_A^2 n_h$.

It is interesting to check the cancellations between the bare four-loop
expression and the mass counterterm contributions (which are known
analytically and can be found in the ancillary file to this
paper~\cite{progdata}). For this reason we show in Tab.~\ref{tab::Z2OSxi0noCT}
the bare four-loop coefficients. The comparison with the corresponding entries
in Tab.~\ref{tab::Z2OSxi0} shows that the coefficients of some of the colour
structures suffer from large cancellations which in some cases is even more
than two orders of magnitude (see, e.g., the $C_F^3n_l$ term). Note that
the numerically dominant colour structure $C_F C_A^3$ is not affected by mass
renormalization.

In Ref.~\cite{Grozin:2017css} the pole of the colour structure
$n_l(d_F^{abcd})^2$ has been determined from the requirement that a
certain combination of renormalization constants in full QCD and HQET
are finite (see also discussion in Section~\ref{sec::HQET} below).
Its analytic expression in our notation reads
\begin{eqnarray}
  \delta Z_2^{d_{FF}L} &=&
 - \frac{1}{\epsilon}\left(\frac{1}{8} + \frac{\pi^2}{12} - \frac{\zeta_3}{8} 
    - \frac{\pi^2\zeta_3}{12} + \frac{5\zeta_5}{32} \right) +\ldots
  \nonumber\\
  &\approx& \frac{0.0294223}{\epsilon} + \ldots
  \,,
  \label{eq::dFFL_pole}
\end{eqnarray}
which has to be compared to our numerical result $(0.011 \pm 0.064)/\epsilon +
\ldots$ (see Tab.~\ref{tab::Z2OSxi0}). The result in
Eq.~(\ref{eq::dFFL_pole}) agrees with our result within the uncertainty. Note,
however, that the absolute value of this contribution is quite small which
explains our large relative uncertainty.

It is interesting to insert the numerical values of the colour factors
and evaluate $\delta Z_2^{(4)}$ for $N_c=3$.  The obtain the
corresponding expression we choose $N_c=3$ after inserting the master
integrals but before combining the uncertainties from the various
$\epsilon$ expansion coefficients of the colour factors. The results
for the various powers of $n_l$ are given in Tab.~\ref{tab::Z2OSxi0nc3}.
Note that for $\xi=0$ (top) all uncertainties are of order $10^{-4}$.
Furthermore, for all powers of $n_l$ we observe non-zero
coefficients in the poles up to fourth order.
For completeness we present in Tab.~\ref{tab::Z2OSxi0nc3} also
results for the $\xi^1$ and $\xi^2$ terms.
For the linear $\xi$ coefficients we observe non-zero entries
only for the $n_l^0$ and the linear-$n_l$ term.
The coefficients of $\xi^2$ are only non-zero for the $n_l^0$
contribution.

\begin{table}[ht]
\begin{center}
{\scalefont{0.7}
  \begin{tabular}{c|c|c|c|c|c}
    \hline
    $\xi=0$ & $1/\epsilon^4$  & $1/\epsilon^3$
    & $1/\epsilon^2$  & $1/\epsilon$ & $\epsilon^0$  \\
    \hline
$n_l^0$ &
$    -1.77242 \pm      0.00040$ &
$    -27.6674 \pm       0.0041$ &
$    -317.093 \pm        0.029$ &
$    -3142.15 \pm         0.33$ &
$    -28709.9 \pm          3.2$ 
\\
$n_l^1$ &
$    0.460936 \pm     0.000016$ &
$     6.69143 \pm      0.00023$ &
$     74.6540 \pm       0.0013$ &
$    696.6612 \pm       0.0076$ &
$    6174.290 \pm        0.084$ 
\\
$n_l^2$ &
$        -0.039931$ &
$         -0.51572$ &
$          -5.5055$ &
$          -48.777$ &
$          -418.93$ 
\\
$n_l^3$ &
$       0.00115741$ &
$        0.0125386$ &
$         0.126757$ &
$          1.07105$ &
$           8.9160$ 
\\

    \hline
  \end{tabular}
  \\[1em]
  \begin{tabular}{c|c|c|c|c|c}
    \hline
    $\xi^1$ & $1/\epsilon^4$  & $1/\epsilon^3$
    & $1/\epsilon^2$  & $1/\epsilon$ & $\epsilon^0$  \\
    \hline
$n_l^0$ &
$   -0.018555 \pm     0.000011$ &
$    0.034239 \pm     0.000089$ &
$    -0.05678 \pm      0.00052$ &
$      5.2230 \pm       0.0028$ &
$      36.820 \pm        0.017$ 
\\
$n_l^1$ &
$       0.00173611$ &
$       -0.0052083$ &
$        0.0224269$ &
$         -0.34863$ &
$         -1.61105$ 
\\

    \hline
  \end{tabular}
  \\[1em]
  \begin{tabular}{c|c|c|c|c|c}
    \hline
    $\xi^2$ & $1/\epsilon^4$  & $1/\epsilon^3$
    & $1/\epsilon^2$  & $1/\epsilon$ & $\epsilon^0$  \\
    \hline
$n_l^0$ &
$   0.0000002 \pm    0.0000038$ &
$    0.001952 \pm     0.000026$ &
$    -0.03022 \pm      0.00012$ &
$    -0.18686 \pm      0.00061$ &
$     -2.9266 \pm       0.0028$ 
\\

    \hline
  \end{tabular}
}
\caption{\label{tab::Z2OSxi0nc3}Results for the coefficients of
  $\delta Z_2^{(4)}$ after choosing $N_c=3$.
  The $\xi=0$, $\xi^1$ and $\xi^2$ contributions
  are shown in the top, middle and bottom table.
  A security factor~10 has been applied to the
  uncertainties.}
\end{center}
\end{table}

Finally, we discuss the wave function renormalization for QED. It is obtained
from the QCD result by adopting the following values for the QCD colour
factors
\begin{eqnarray}
  C_F \to 1\,,\quad C_A\to0\,,\quad T\to1\,,\quad
  d_F^{abcd}\to 1\,,\quad d_A^{abcd}\to0\,,\quad N_c\to 1
  \,.
\end{eqnarray}
We furthermore set $n_h=1$ but keep the dependence on $n_l$.  Note
that $n_l=0$ corresponds to the case of a massive electron and $n_l=1$
describes the case of a massive muon and a massless electron.
Our results are shown in Tab.~\ref{tab::QED}.
For the $n_l$-independent part we have an uncertainty of about 10\%,
the $n_l^1$ term is determined with a 3\% accuracy.

\begin{table}[t]
\begin{center}
{\scalefont{0.75}
  \begin{tabular}{c|c|c|c|c|c}
    \hline
    & $1/\epsilon^4$  & $1/\epsilon^3$
    & $1/\epsilon^2$  & $1/\epsilon$ & $\epsilon^0$  \\
    \hline
$n_l^0$ &
$     0.20500 \pm      0.00037$ &
$      0.5980 \pm       0.0027$ &
$      -0.895 \pm        0.021$ &
$       -6.18 \pm         0.17$ &
$       -17.4 \pm          1.6$ 
\\
$n_l^1$ &
$     0.17058 \pm      0.00011$ &
$      0.9556 \pm       0.0014$ &
$      2.9397 \pm       0.0079$ &
$      10.480 \pm        0.064$ &
$       25.92 \pm         0.80$ 
\\
$n_l^2$ &
$         0.056424$ &
$          0.46123$ &
$          3.03509$ &
$          18.7456$ &
$          105.069$ 
\\
$n_l^3$ &
$        0.0069444$ &
$         0.075231$ &
$          0.76054$ &
$           6.4263$ &
$           53.496$ 
\\

    \hline
  \end{tabular}
}
   \caption{\label{tab::QED}Results for $Z_2^{\rm OS}$ specified to QED.}
\end{center}
\end{table}

The on-shell wave function renormalization constant in QED has to be
independent of $\xi$~\cite{Johnson:1959zz,Melnikov:2000zc}
which is fulfilled
in our result as can be seen from the absence of all abelian
coefficients in Tabs.~\ref{tab::Z2OSxi1} and~\ref{tab::Z2OSxi2}; they
are analytically zero.
Note that the gauge parameter dependence only cancels after adding the mass counterterm
contributions.


\newcommand{\ZHQET}{\ensuremath{Z_2^{\mathrm{HQET}}}}
\newcommand{\ZMS}{\ensuremath{Z_2^{\overline{\mathrm{MS}}}}}
\newcommand{\ZOS}{\ensuremath{Z_2^{\mathrm{OS}}}}
\newcommand{\api}{\ensuremath{\left(\frac{\alpha_s}{\pi}\right)}}
\newcommand{\afpi}{\ensuremath{\left(\frac{\alpha_s}{4\pi}\right)}}

\section{\label{sec::HQET}Checks and HQET wave function renormalization}

In this Section we describe several checks of our results.  In particular, we
discuss the relation to the wave function renormalization constant in HQET.

We start with the discussion of the $\overline{\rm MS}$ wave function
renormalization constant $\ZMS$ which has been obtained to five-loop accuracy
in Refs.~\cite{Luthe:2016xec,Baikov:2017ujl}. In these papers also the
full $\xi$-dependence at four loops has been computed which is crucial for our
application. By definition it only contains
ultraviolet poles. On the other hand, as discussed in the Introduction,
$\ZOS$, contains both ultraviolet and infrared poles since it has to take care
of both types of divergences in processes containing external heavy
quarks. The ultraviolet divergences of $\ZOS$ have to agree with the ones of
$\ZMS$ and thus $\ZMS/\ZOS$ only contains infrared poles.  Note that the
latter have to agree with the ultraviolet poles of the wave function
renormalization constant in HQET, $\ZHQET$, which can be seen as follows (see
also discussion in Ref.~\cite{Melnikov:2000zc}): The off-shell heavy-quark
propagator is infrared finite and contains only ultraviolet divergences, which
can be renormalized in the $\overline{\rm MS}$ scheme, i.e., they are taken
care of by $\ZMS$. If one applies an asymptotic
expansion~\cite{Beneke:1997zp,Smirnov:2012gma} around the on-shell limit one
obtains two contributions. The first one corresponds to a naive Taylor expansion of
on-shell integrals which have to be evaluated in full QCD. It develops both
ultraviolet and infrared divergences, as discussed above for the case of
$\ZOS$.  The second contribution corresponds to HQET integrals and only has
ultraviolet poles which have to cancel the infrared poles of the QCD
contribution. Note that the wave functions $\ZOS$ and $\ZHQET$ considered in
this paper correspond to the leading term in the expansion and thus
$\ZHQET/\ZOS$ has to be infrared finite. As a consequence, the following
combination of renormalization constants
\begin{equation}
  \frac{\ZMS}{\ZOS}\ZHQET
  \label{eq::hqetfinite}
\end{equation}
has to be finite (see also discussion in Ref.~\cite{Grozin:2010wa}). 
We will use this fact to determine the poles of
$\ZHQET$.

HQET describes the limit of QCD where the mass of the heavy quark goes to
infinity.  The heavy quark field is integrated out from the Lagrange
density. Thus, it is not a dynamical degree of freedom any more. As a
consequence, HQET contains as parameters the strong coupling constant and
gauge parameter defined in the $n_l$-flavour theory, $\alpha_s^{(n_l)}$ and
$\xi^{(n_l)}$.\footnote{Note that all quantities discussed in
  Section~\ref{sec::Z2OS} depend on $n_f=n_l+n_h$ flavours.}  Furthermore, there
are no closed heavy quark loops, i.e., colour factors involving $n_h$ are
absent.  Thus, when constructing (\ref{eq::hqetfinite}) we can check that
in the ratio $\ZMS/\ZOS$ all colour structures containing $n_h$ are finite
after using the decoupling relations for $\alpha_s$ and
$\xi$~\cite{Chetyrkin:1997un}.
At two- and three-loop order this check can be performed analytically. At four
loops we observe that $\ZMS/\ZOS$ is finite within our numerical precision.
Note that this concerns the eleven colour structures in
Eq.~(\ref{eq::z24_col}) which are proportional to $n_h$, $n_h^2$ or $n_h^3$.
Let us mention that all coefficients are zero within three standard deviations
of the original {\tt FIESTA} uncertainty which means that in this case a
security factor~3 would be sufficient.

The remaining twelve four-loop colour structures are present in $\ZHQET$
and the corresponding pole term can be extracted from
Eq.~(\ref{eq::hqetfinite}).
Before presenting the results we remark that
$\ZHQET$ exponentiates according to~\cite{Melnikov:2000zc,Chetyrkin:2003vi}
\begin{eqnarray}
  \ZHQET &=& \exp \left\{ x_1 C_F \api 
    + C_F \left[ x_2 C_A + x_3 T n_l \right] \api^2 
    + C_F \left[ x_4 C_A ^2 + x_5 C_A T n_l
  \right.\right.\nonumber \\&& \left.\left.
    + x_6 T^2 n_l^2 + x_7 C_F T n_l \right] \api^3
    + \left[C_F \left(
    x_8 C_A^3 
    + x_9 C_A^2 T n_l 
    + x_{10} C_A T^2 n_l^2 
    \right.\right.\right.\nonumber \\&& \left.\left.\left.
    + x_{11} T^3 n_l^3
    + x_{12} C_F^2 T n_l
    + x_{13} C_F C_A T n_l
    + x_{14} C_F T^2 n_l^2\right)
    \right.\right.\nonumber \\&& \left.\left.
    + x_{15} d_F^{abcd}d_A^{abcd}/N_c
    + x_{16} d_F^{abcd}d_F^{abcd}n_l/N_c
  \right] \api^4
 + \ldots \right\}\,,
  \label{eq:HQETexp}
\end{eqnarray}
and thus there are only nine genuinely new colour coefficients at four loops
($x_8,\ldots,x_{16}$) and the remaining three contributions proportional to
$C_F^4, C_F^3 C_A$ and $C_F^2 C_A^2$ can be predicted from lower loop
orders. The comparison with the explicit calculation provides a strong check on our
calculation.  Note that the predictions of the $C_F^4, C_F^3 C_A$ and $C_F^2
C_A^2$ contributions are available in analytic form.

In our practical calculations we proceed as follows: In a first step we use
Eq.~(\ref{eq::hqetfinite}) to obtain a result for $\ZHQET$ from the
requirement that the combination of the three quantities is finite.  Afterwards
we use this result and compare to the expanded version of
Eq.~(\ref{eq:HQETexp}) to determine the coefficients $x_i$.  Finally, we use
Eq.~(\ref{eq:HQETexp}) to predict the $C_F^4, C_F^3 C_A$ and $C_F^2 C_A^2$ of
$\ZHQET$.

We refrain from providing explicit results for
$\ZHQET$ but provide our results for $x_i$ in the ancillary file to this
paper~\cite{progdata}. Furthermore, we present the expressions
for the corresponding anomalous dimension, which is  given by
\begin{eqnarray}
  \gamma_{\rm HQET} &=& \frac{{\rm d}\log \ZHQET}{{\rm d}\log \mu^2}
  \nonumber\\
  &=& \sum_{n\ge1} \gamma_{\rm HQET}^{(n)} \left(\frac{\alpha_s(\mu)}{\pi}\right)^n
  \,.
\end{eqnarray}
Since our four-loop expression for $\ZHQET$ is only known
numerically, we have spurious $\epsilon$ poles in $\gamma_{\rm HQET}$.
However, all of them are zero within two standard deviations
of the uncertainty provided by {\tt FIESTA},
which constitutes another useful cross check for our calculation.

Let us in the following present our results for $\gamma_{\rm HQET}$.
Up to three-loop order we have
\begin{eqnarray}
  \gamma_{\rm HQET}^{(1)} &=& -\frac{\cR}{2}\left(1 + \frac{\xinl}{2} \right)
  \,, \nonumber\\
  \gamma_{\rm HQET}^{(2)} &=& 
  \cR\cA\left( -\frac{19}{24} - \frac{5\xinl}{32} + \frac{(\xinl)^2}{64}\right)
  + \frac{\cR\tr n_l}{3}
  \,, \nonumber\\
  \gamma_{\rm HQET}^{(3)} &=& 
  \cR\cA^2 \left[-\frac{19495}{27648} 
  - \frac{3\zeta_3}{16}
  - \frac{\pi^4}{360}   
  + \xinl \left(- \frac{379}{2048} - \frac{15\zeta_3}{256}+ \frac{\pi^4}{1440} \right)
  \right.\nonumber\\&&\left.\mbox{}
  + (\xinl)^2\left( \frac{69}{2048} + \frac{3\zeta_3}{512}\right)
  - \frac{5(\xinl)^3}{1024}
  \right]
  + \cR\cA\tr n_l \left(\frac{1105}{6912} 
    + \frac{3\zeta_3}{4}
    + \frac{17\xinl}{256} 
  \right)
  \nonumber\\&&\mbox{}
  + \cR^2\tr n_l \left(\frac{51}{64} -\frac{3\zeta_3}{4}\right)
  + \frac{5\cR\tr^2 n_l^2}{108}
  \,,
  \label{eq::gammaHQET}
\end{eqnarray}
which agree with Refs.~\cite{Melnikov:2000zc,Chetyrkin:2003vi}.

The four-loop terms to $\gamma_{\rm HQET}$ can be found in
Tab.~\ref{tab::gammaHQET4} where for each colour factor the coefficients of
the $(\xinl)^k$ terms are shown together with their uncertainty. As for $\ZOS$
in Section~\ref{sec::Z2OS} we have introduced a security factor~10.  Note that
the coefficients of $(\xinl)^k$ with $k\ge3$ have not been computed.

\begin{table}[t]
  \centering
  \begin{tabular}{c|c|c|c}
    & \mbox{}\hfill$(\xinl)^0$\hfill\mbox{} 
    & \mbox{}\hfill$(\xinl)^1$\hfill\mbox{} 
    & \mbox{}\hfill$(\xinl)^2$\hfill\mbox{} \\
    \hline
$FAAA$ &$     {-2.03} \pm 0.35$ &$  -0.29037 \pm      0.00052$ &$    {0.07083} \pm      0.00010$\\
$d_{FA}$ &$    {1.53} \pm 0.84$ &$  {0.5083} \pm       0.0098$ &$   {-0.1031} \pm       0.0024$\\
$d_{FF}L$ &$     {0.54} \pm         0.26$ &  & \\
$FFFL$ &$    {0.1894} \pm       0.0030$ &  & \\
$FFAL$ &$    {-0.4566} \pm       0.0055$ &$  -0.0076630$ & \\
$FAAL$ &$      {2.576} \pm       {0.010}$ &$     0.25147$ &$  -0.0103348$\\
$FFLL$ &$     0.25725$ &  & \\
$FALL$ &$    -0.53745$ &$  -0.0077460$ & \\
$FLLL$ &$   -0.048262$ &  & \\
    \hline
  \end{tabular}
  \caption{\label{tab::gammaHQET4}Results for the different 
    colour factors of $\gamma_{\rm HQET}^{(4)}$. In the columns two to four
    the coefficients of different powers of $\xinl$ are given.
    In the uncertainties a security factor~10 has been introduced.}
\end{table}

We have the worst precision of about 50\% for the colour factors
$d_F^{abcd}d_A^{abcd}$ and $n_l d_F^{abcd}d_F^{abcd}$ followed by $C_F C_A^3$
which is 17\%. The relative uncertainty of the remaining $n_l$ terms is much
smaller. Note that the $n_l^2$ and $n_l^3$ terms are known analytically.
They are obtained in a straightforward way for the corresponding analytic
results for $\ZOS$ from Ref.~\cite{Lee:2013sx}. Our results read
\begin{eqnarray}
  \gamma_{\rm HQET}^{(4),FFLL} &=&
  \frac{3 \zeta _3}{4}-\frac{\pi ^4}{240}-\frac{103}{432}
  \,,\nonumber\\
  \gamma_{\rm HQET}^{(4),FALL} &=&
  -\frac{35 \zeta _3}{48}
  +\frac{\pi^4}{240}-\frac{4157}{62208}
  +\xinl\left(-\frac{\zeta _3 }{48}+\frac{269 }{15552}\right)
  \,,\nonumber\\
  \gamma_{\rm HQET}^{(4),FLLL} &=&
  \frac{1}{54}-\frac{\zeta _3}{18} 
  \,.
\end{eqnarray}
The expression for $\gamma_{\rm HQET}^{(4),FFLL}$ agrees with
Ref.~\cite{Grozin:2015kna,Grozin:2016ydd} and $\gamma_{\rm HQET}^{(4),FLLL}$
can be found in Ref.~\cite{Broadhurst:1994se}. $\gamma_{\rm HQET}^{(4),FALL}$
is new.

Recently also for the $n_l d_F^{abcd}d_F^{abcd}$ colour structure 
analytic results have been obtained~\cite{Grozin:2017css}. Their result reads
\begin{eqnarray}
  \gamma_{\rm HQET}^{(4),d_{FF}L} &=&     
  -\frac{5}{8} \zeta_5 + \frac{1}{3} \pi^2 \zeta_3 
  + \frac{1}{2} \zeta_3 - \frac{1}{3} \pi^2  \approx 0.617689\ldots
  \,,
\end{eqnarray}
and agrees well with our findings $\gamma_{\rm HQET}^{(4),d_{FF}L} \approx
0.54\pm 0.26$. Note that here a security factor~2 would have been
sufficient.

There are no contributions from the colour structures $C_F^4, C_F^3 C_A$ and
$C_F^2 C_A^2$ to $\gamma_{\rm HQET}^{(4)}$ as it is obvious by inspecting
Eq.~(\ref{eq:HQETexp}): the four-loop $C_F^4, C_F^3 C_A$ and $C_F^2 C_A^2$
terms are generated by products of lower-order contributions. Since all
coefficients $x_i$ only contain poles in $\epsilon$, the $1/\epsilon$ pole of
$\ZHQET$ does not involve $C_F^4, C_F^3 C_A$ and $C_F^2 C_A^2$.

Let us finally compare the predicted $C_F^4, C_F^3 C_A$ and $C_F^2 C_A^2$
contributions to $\ZHQET$ to the ones we obtain by an explicit calculation.
Tab.~\ref{tab::Z2HQET_cR} contains coefficients of $(\xinl)^k \epsilon^n$ for
$k=0,1$ and $2$ and for values of $n=-4,-3,\ldots$ up to one unit higher than the
order up to which the corresponding colour structure has a non-zero
contribution. The last $\epsilon$ order is shown as a check and demonstrates
how well we can reproduce the 0. Note that in this table the displayed uncertainties are not
multiplied by a security factor but correspond to the quadratically combined
FIESTA uncertainties. In some case the relative uncertainty is very small and
thus not shown at all.  In all cases shown in Fig.~\ref{tab::Z2HQET_cR} the
numerical results agree within 1.5 sigma with the analytic predictions from
Eq.~(\ref{eq:HQETexp}).
Note the colour factors $C_F^4, C_F^3 C_A$ and $C_F^2 C_A^2$ get contributions
from the most complicated master integrals and thus
the above comparison provides a strong check on the numerical setup of our
calculation.

\begin{table}[t]
  \centering
  \begin{tabular}{c|l|l|l}
    & \mbox{}\hfill$(\xinl)^0$\hfill\mbox{} 
    & \mbox{}\hfill$(\xinl)^1$\hfill\mbox{} 
    & \mbox{}\hfill$(\xinl)^2$\hfill\mbox{} \\
    \hline
    $C_F^4$ &&& \\
$1/\epsilon^4$
 &$    0.0026042$ &$    0.0052083$ &$    0.0039063$\\
 &$   0.0025932 \pm     0.000025$ &$   0.0052083$ &$   0.0039063$
\\
$1/\epsilon^3$
 &$      0.00000$ &$      0.00000$ &$      0.00000$\\
 &$  0.00013049 \pm      0.00019$ &$     0.00000$ &$     0.00000$
\\
    \hline
    $C_F^3C_A$ &&& \\
$1/\epsilon^4$
 &$     0.035156$ &$     0.044922$ &$     0.016602$\\
 &$    0.035190 \pm      0.00005$ &$    0.044922 $ &$    0.016602$
\\
$1/\epsilon^3$
 &$    -0.049479$ &$    -0.059245$ &$    -0.021159$\\
 &$   -0.049878 \pm      0.00044$ &$   -0.059245 \pm   0.00000006$ &$   -0.021159$
\\
$1/\epsilon^2$
 &$      0.00000$ &$      0.00000$ &$      0.00000$\\
 &$   0.0029893 \pm       0.0041$ &{$-0.0000002 \pm     0.0000020$} &$ 0.00000$
\\
    \hline
    $C_F^2C_A^2$ &&& \\
$1/\epsilon^4$
 &$     0.130914$ &$     0.085558$ &$    0.0027262$\\
 &$    0.130887 \pm      0.00004$ &$    0.085558 \pm   0.00000002$ &$   0.0027262$
\\
$1/\epsilon^3$
 &$     -0.31170$ &$    -0.191497$ &$   -0.0081380$\\
 &$    -0.31133 \pm      0.00035$ &$   -0.191497 \pm    0.0000002$ &$  -0.0081380$
\\
$1/\epsilon^2$
 &$      0.27852$ &$     0.162322$ &$    0.0088241$\\
 &$     0.27669 \pm       0.0033$ &$    0.162323 \pm     0.000002$ &$   0.0088241$
\\
$1/\epsilon^1$
 &$      0.00000$ &$      0.00000$ &$      0.00000$\\
 &$    {0.046} \pm        0.031$ &${-0.000014} \pm     0.000022$ &$0.00000$
\\

  \end{tabular}
  \caption{\label{tab::Z2HQET_cR}Contributions of the colour structures $C_F^4$,
    $C_F^3C_A$ and $C_F^2C_A^2$ to \ZHQET. The coefficients of $(\xinl)^0$,
    $(\xinl)^1$ and $(\xinl)^2$ are given in the rows two to four. For each power of $\epsilon$
    the first row corresponds to the numerical evaluation of the analytic
    result and the second row to the numerical result of our explicit
    calculation of $\ZOS$. Relative uncertainties below $10^{-5}$ are set to
    zero.
    Note that the uncertainties in this paper are not multiplied by a
      security factor~10.}
\end{table}


\section{\label{sec::concl}Conclusions}

We have computed four-loop QCD corrections to the wave function
renormalization constant of heavy quarks, $Z_2^{\rm OS}$. Besides the on-shell
quark mass renormalization constant and the leptonic anomalous magnetic
moment, which have been considered in Refs.~\cite{Marquard:2015qpa,Marquard:2016dcn}
and~\cite{Marquard:2017iib}, respectively, this constitutes a third
``classical'' application of four-loop on-shell integrals.  In the present
calculation we could largely profit from the previous calculations. However,
we had to deal with a more involved reduction to master
integrals. Furthermore, we observed higher $\epsilon$ poles in the prefactors
of some of the master integrals which forced us to either change the basis or
to expand the corresponding master integrals to higher order in $\epsilon$.

$Z_2^{\rm OS}$ is neither gauge parameter independent nor 
infrared finite which excludes two important checks used for
$Z_m^{\rm OS}$ and the anomalous magnetic moment. However,
a number of cross checks are provided by the relation to the
wave function renormalization constant of HQET.

In physical applications $Z_2^{\rm OS}$ enters, among other quantities,
as building block. Most likely in the evaluation of the other pieces
numerical methods play an important role as well and thus
various numerical pieces have to be combined to arrive at physical
cross sections or decay rates. It might be that numerical
cancellations take place and thus, to date, it is not clear whether
the numerical precision reached for $Z_2^{\rm OS}$
(which is of the order of $10^{-4}$ for $N_c=3$)
is sufficient for phenomenological applications. However, the results
obtained in this paper serve for sure as important cross checks for
future more precise or even analytic calculations.

In future, it would, of course, be desireable to obtain analytic results for
fundamental quantities like on-shell QCD renormalization constants as $\ZOS$,
which is considered in this paper, and $Z_m^{\rm OS}$ from
Refs.~\cite{Marquard:2015qpa,Marquard:2016dcn}. First steps in this direction
have been undertaken in Ref.~\cite{Laporta:2017okg} where a semi-analytic
approach has been used to obtain a high-precision result for the anomalous
magnetic moment of the electron. One could imagine to
extend this anlysis to the QCD-like master integrals.


\section*{Acknowledgements}

The work of A.S. and V.S. is supported by RFBR, grant 17-02-00175A.
We thank the High Performance Computing Center Stuttgart (HLRS) for
providing computing time used for the numerical computations with {\tt
  FIESTA}.  The research is carried out using the equipment of the
shared research facilities of HPC computing resources at Lomonosov
Moscow State University.  P.M was supported in part by the EU Network
HIGGSTOOLS PITN-GA-2012-316704.
We thank Andrey Grozin for carefully reading the manuscript and many
useful comments.


\begin{appendix}


\section{\label{app::tables}Numerical results for $\ZOS$}

Tables~\ref{tab::Z2OSxi0},~\ref{tab::Z2OSxi1},~\ref{tab::Z2OSxi2}
and~\ref{tab::Z2OSxi0noCT} contain the numerical results
for the coefficients of the individual colour factors contributing to $\ZOS$.

\begin{sidewaystable}[ht]
  \centering
      {\scalefont{0.8}
        \begin{tabular}{c|c|c|c|c|c}
          & $1/\epsilon^4$  & $1/\epsilon^3$
          & $1/\epsilon^2$  & $1/\epsilon$ & $\epsilon^0$  \\
          \hline
$FFFF$ &
$     0.01317 \pm      0.00025$ &
$      0.0836 \pm       0.0019$ &
$      -0.084 \pm        0.017$ &
$       -1.96 \pm         0.16$ &
$        -4.1 \pm          1.5$ 
\\
$FFFA$ &
$    -0.09665 \pm      0.00053$ &
$     -0.7611 \pm       0.0044$ &
$      -1.275 \pm        0.041$ &
$        1.10 \pm         0.38$ &
$        -9.8 \pm          3.6$ 
\\
$FFAA$ &
$     0.21661 \pm      0.00040$ &
$      2.1150 \pm       0.0035$ &
$       9.698 \pm        0.033$ &
$       57.52 \pm         0.31$ &
$       324.5 \pm          2.9$ 
\\
$FAAA$ &
$    -0.14442 \pm      0.00011$ &
$    -1.76642 \pm      0.00096$ &
$    -14.4491 \pm       0.0092$ &
$    -123.354 \pm        0.086$ &
$    -1007.40 \pm         0.82$ 
\\
$d_{FA}$ &
$    -0.00002 \pm      0.00029$ &
$      0.0006 \pm       0.0033$ &
$      -0.002 \pm        0.024$ &
$        0.40 \pm         0.21$ &
$         9.4 \pm          2.1$ 
\\
$d_{FF}L$ &
$     0.00001 \pm      0.00011$ &
$     -0.0001 \pm       0.0014$ &
$      0.0000 \pm       0.0079$ &
$       0.011 \pm        0.064$ &
$       -2.18 \pm         0.80$ 
\\
$d_{FF}H$ &
$    -0.00001 \pm      0.00023$ &
$      0.0001 \pm       0.0015$ &
$      -0.001 \pm        0.011$ &
$      -0.120 \pm        0.076$ &
$        0.10 \pm         0.50$ 
\\
$FFFL$ &
$   0.0351561 \pm    0.0000013$ &
$   0.2499987 \pm    0.0000092$ &
$    0.496651 \pm     0.000077$ &
$     0.39174 \pm      0.00074$ &
$      1.3920 \pm       0.0067$ 
\\
$FFAL$ &
$  -0.1575519 \pm    0.0000033$ &
$   -1.457029 \pm     0.000022$ &
$    -7.60181 \pm      0.00016$ &
$    -46.0162 \pm       0.0014$ &
$    -236.417 \pm        0.012$ 
\\
$FAAL$ &
$   0.1575515 \pm    0.0000052$ &
$    1.889980 \pm     0.000070$ &
$    17.10515 \pm      0.00039$ &
$    145.3220 \pm       0.0026$ &
$    1190.195 \pm        0.031$ 
\\
$FFLL$ &
$        0.0286458$ &
$         0.244792$ &
$          1.37840$ &
$           8.3824$ &
$           40.329$ 
\\
$FALL$ &
$        -0.057292$ &
$         -0.66059$ &
$          -6.3943$ &
$          -54.229$ &
$          -447.65$ 
\\
$FLLL$ &
$        0.0069444$ &
$         0.075231$ &
$          0.76054$ &
$           6.4263$ &
$           53.496$ 
\\
$FFFH$ &
$    0.070313 \pm     0.000023$ &
$    0.255860 \pm     0.000093$ &
$    -0.65497 \pm      0.00055$ &
$     -3.8002 \pm       0.0036$ &
$      -5.953 \pm        0.019$ 
\\
$FFAH$ &
$    -0.26173 \pm      0.00010$ &
$    -1.58102 \pm      0.00044$ &
$     -3.2136 \pm       0.0021$ &
$     -11.729 \pm        0.013$ &
$     -26.860 \pm        0.083$ 
\\
$FAAH$ &
$    0.215336 \pm     0.000061$ &
$     1.95402 \pm      0.00027$ &
$     11.5396 \pm       0.0014$ &
$     70.3186 \pm       0.0091$ &
$     424.301 \pm        0.056$ 
\\
$FFHH$ &
$   0.0937498 \pm    0.0000014$ &
$   0.2109378 \pm    0.0000059$ &
$   -0.329095 \pm     0.000035$ &
$    -0.57438 \pm      0.00013$ &
$    -7.99681 \pm      0.00079$ 
\\
$FAHH$ &
$   -0.117186 \pm     0.000011$ &
$   -0.681863 \pm     0.000054$ &
$    -2.52735 \pm      0.00029$ &
$    -10.3208 \pm       0.0012$ &
$    -40.2646 \pm       0.0062$ 
\\
$FHHH$ &
$        0.0277778$ &
$         0.047454$ &
$         0.173582$ &
$         0.276902$ &
$          0.61212$ 
\\
$FFLH$ &
$         0.093750$ &
$          0.50781$ &
$   1.3923245 \pm    0.0000012$ &
$    5.834231 \pm     0.000010$ &
$    8.990228 \pm     0.000074$ 
\\
$FALH$ &
$  -0.1545138 \pm    0.0000011$ &
$  -1.3179979 \pm    0.0000063$ &
$   -9.088033 \pm     0.000034$ &
$   -56.32679 \pm      0.00020$ &
$   -344.7315 \pm       0.0015$ 
\\
$FLLH$ &
$        0.0277778$ &
$         0.216435$ &
$          1.65669$ &
$          10.3632$ &
$           64.740$ 
\\
$FLHH$ &
$         0.041667$ &
$         0.197917$ &
$          1.05074$ &
$           4.2433$ &
$          17.7160$ 
\\
        \end{tabular}
      }
      \caption{\label{tab::Z2OSxi0}Results for the coefficients of
        $\delta Z_2^{(4)}$ as defined in Eq.~(\ref{eq::z24_col}) for
        $\xi=0$. A security factor~10 has been applied to the
        uncertainties.}
\end{sidewaystable}

\begin{sidewaystable}[ht]
  \centering
      {\scalefont{0.8}
        \begin{tabular}{c|c|c|c|c|c}
          & $1/\epsilon^4$  & $1/\epsilon^3$
          & $1/\epsilon^2$  & $1/\epsilon$ & $\epsilon^0$  \\
          \hline
$FFFA$ &
$     0$ &
$     0$ &
$   -0.000000 \pm     0.000020$ &
$     0.00001 \pm      0.00020$ &
$     -0.0001 \pm       0.0011$ 
\\
$FFAA$ &
$     0$ &
$  -0.0000001 \pm    0.0000016$ &
$   -0.005369 \pm     0.000022$ &
$    -0.03679 \pm      0.00022$ &
$     -0.3166 \pm       0.0012$ 
\\
$FAAA$ &
$     0$ &
$  -0.0000001 \pm    0.0000039$ &
$    0.013200 \pm     0.000023$ &
$     0.11976 \pm      0.00013$ &
$     1.42164 \pm      0.00076$ 
\\
$d_{FA}$ &
$  -0.0000003 \pm    0.0000100$ &
$    0.000005 \pm     0.000088$ &
$    -0.00000 \pm      0.00051$ &
$      0.1135 \pm       0.0025$ &
$       0.147 \pm        0.013$ 
\\
$FAAL$ &
$     0$ &
$     0$ &
$       -0.0035799$ &
$        -0.033281$ &
$         -0.40121$ 
\\
$FFAH$ &
$        0.0039062$ &
$       -0.0094401$ &
$   0.0069760 \pm    0.0000032$ &
$    0.037345 \pm     0.000035$ &
$    -0.76089 \pm      0.00024$ 
\\
$FAAH$ &
$       -0.0052626$ &
$        0.0112034$ &
$  -0.0854353 \pm    0.0000018$ &
$    0.216644 \pm     0.000018$ &
$    -1.40360 \pm      0.00013$ 
\\
$FAHH$ &
$       0.00260417$ &
$       -0.0078125$ &
$         0.047919$ &
$        -0.182917$ &
$          0.78980$ 
\\
$FALH$ &
$       0.00173611$ &
$       -0.0052083$ &
$         0.043906$ &
$        -0.148948$ &
$          0.79619$ 
\\

        \end{tabular}
      }
      \caption{\label{tab::Z2OSxi1}Same as in Tab.~\ref{tab::Z2OSxi0}
        but the coefficients of the linear $\xi$ terms.}
\end{sidewaystable}

\begin{sidewaystable}[ht]
  \centering
      {\scalefont{0.8}
        \begin{tabular}{c|c|c|c|c|c}
          & $1/\epsilon^4$  & $1/\epsilon^3$
          & $1/\epsilon^2$  & $1/\epsilon$ & $\epsilon^0$  \\
          \hline
$FAAA$ &
$     0$ &
$   0.0000000 \pm    0.0000011$ &
$  -0.0006711 \pm    0.0000052$ &
$   -0.005817 \pm     0.000025$ &
$    -0.07062 \pm      0.00012$ 
\\
$d_{FA}$ &
$   0.0000002 \pm    0.0000038$ &
$   -0.000001 \pm     0.000026$ &
$    -0.00000 \pm      0.00012$ &
$    -0.01250 \pm      0.00061$ &
$     -0.0748 \pm       0.0028$ 
\\
$FAAH$ &
$     0$ &
$       0.00032552$ &
$      -0.00100945$ &
$        0.0089715$ &
$        -0.032896$ 
\\
        \end{tabular}
      }
      \caption{\label{tab::Z2OSxi2}Same as in Tab.~\ref{tab::Z2OSxi0}
        but the coefficients of $\xi^2$.}
\end{sidewaystable}

\begin{sidewaystable}[ht]
  \centering
      {\scalefont{0.8}
        \begin{tabular}{c|c|c|c|c|c}
          & $1/\epsilon^4$  & $1/\epsilon^3$
          & $1/\epsilon^2$  & $1/\epsilon$ & $\epsilon^0$  \\
          \hline
$FFFF$ &
$    -2.73194 \pm      0.00025$ &
$      1.9450 \pm       0.0019$ &
$     -22.265 \pm        0.017$ &
$       93.41 \pm         0.16$ &
$       167.9 \pm          1.5$ 
\\
$FFFA$ &
$    -2.62790 \pm      0.00053$ &
$     -4.6110 \pm       0.0044$ &
$     -48.022 \pm        0.041$ &
$      -74.51 \pm         0.38$ &
$      -618.5 \pm          3.6$ 
\\
$FFAA$ &
$    -1.02981 \pm      0.00040$ &
$     -5.9684 \pm       0.0035$ &
$     -44.525 \pm        0.033$ &
$     -255.18 \pm         0.31$ &
$     -1664.9 \pm          2.9$ 
\\
$FAAA$ &
$    -0.14442 \pm      0.00011$ &
$    -1.76642 \pm      0.00096$ &
$    -14.4491 \pm       0.0092$ &
$    -123.354 \pm        0.086$ &
$    -1007.40 \pm         0.82$ 
\\
$d_{FA}$ &
$    -0.00002 \pm      0.00029$ &
$      0.0006 \pm       0.0033$ &
$      -0.002 \pm        0.024$ &
$        0.40 \pm         0.21$ &
$         9.4 \pm          2.1$ 
\\
$d_{FF}L$ &
$     0.00001 \pm      0.00011$ &
$     -0.0001 \pm       0.0014$ &
$      0.0000 \pm       0.0079$ &
$       0.011 \pm        0.064$ &
$       -2.18 \pm         0.80$ 
\\
$d_{FF}H$ &
$    -0.00001 \pm      0.00023$ &
$      0.0001 \pm       0.0015$ &
$      -0.001 \pm        0.011$ &
$      -0.120 \pm        0.076$ &
$        0.10 \pm         0.50$ 
\\
$FFFL$ &
$   1.3124999 \pm    0.0000013$ &
$   2.6503893 \pm    0.0000092$ &
$   31.883671 \pm     0.000077$ &
$    55.08300 \pm      0.00074$ &
$    477.6787 \pm       0.0067$ 
\\
$FFAL$ &
$   0.8750002 \pm    0.0000033$ &
$    4.729724 \pm     0.000022$ &
$    42.82731 \pm      0.00016$ &
$    217.1492 \pm       0.0014$ &
$    1568.153 \pm        0.012$ 
\\
$FAAL$ &
$   0.1575515 \pm    0.0000052$ &
$    1.889980 \pm     0.000070$ &
$    17.10515 \pm      0.00039$ &
$    145.3220 \pm       0.0026$ &
$    1190.195 \pm        0.031$ 
\\
$FFLL$ &
$        -0.179688$ &
$         -0.88281$ &
$          -9.3076$ &
$          -43.714$ &
$          -340.67$ 
\\
$FALL$ &
$        -0.057292$ &
$         -0.66059$ &
$          -6.3943$ &
$          -54.229$ &
$          -447.65$ 
\\
$FLLL$ &
$        0.0069444$ &
$         0.075231$ &
$          0.76054$ &
$           6.4263$ &
$           53.496$ 
\\
$FFFH$ &
$    2.156250 \pm     0.000023$ &
$    1.492579 \pm     0.000093$ &
$    20.69169 \pm      0.00055$ &
$      9.3957 \pm       0.0036$ &
$      92.254 \pm        0.019$ 
\\
$FFAH$ &
$     1.22916 \pm      0.00010$ &
$     4.77852 \pm      0.00044$ &
$     26.4927 \pm       0.0021$ &
$     101.634 \pm        0.013$ &
$     487.440 \pm        0.083$ 
\\
$FAAH$ &
$    0.215336 \pm     0.000061$ &
$     1.95402 \pm      0.00027$ &
$     11.5396 \pm       0.0014$ &
$     70.3186 \pm       0.0091$ &
$     424.301 \pm        0.056$ 
\\
$FFHH$ &
$  -0.4166669 \pm    0.0000014$ &
$  -0.6484372 \pm    0.0000059$ &
$   -3.496296 \pm     0.000035$ &
$    -7.42824 \pm      0.00013$ &
$   -20.15261 \pm      0.00079$ 
\\
$FAHH$ &
$   -0.117186 \pm     0.000011$ &
$   -0.681863 \pm     0.000054$ &
$    -2.52735 \pm      0.00029$ &
$    -10.3208 \pm       0.0012$ &
$    -40.2646 \pm       0.0062$ 
\\
$FHHH$ &
$        0.0277778$ &
$         0.047454$ &
$         0.173582$ &
$         0.276902$ &
$          0.61212$ 
\\
$FFLH$ &
$         -0.50521$ &
$         -1.68750$ &
$ -13.1485794 \pm    0.0000012$ &
$  -44.062536 \pm     0.000010$ &
$ -243.607004 \pm     0.000074$ 
\\
$FALH$ &
$  -0.1545138 \pm    0.0000011$ &
$  -1.3179979 \pm    0.0000063$ &
$   -9.088033 \pm     0.000034$ &
$   -56.32679 \pm      0.00020$ &
$   -344.7315 \pm       0.0015$ 
\\
$FLLH$ &
$        0.0277778$ &
$         0.216435$ &
$          1.65669$ &
$          10.3632$ &
$           64.740$ 
\\
$FLHH$ &
$         0.041667$ &
$         0.197917$ &
$          1.05074$ &
$           4.2433$ &
$          17.7160$ 
\\

        \end{tabular}
      }
      \caption{\label{tab::Z2OSxi0noCT}Results for the coefficients of
        $\delta Z_2^{(4)}$ as defined in Eq.~(\ref{eq::z24_col}) for
        $\xi=0$ and without taking into account the mass counterterms
        from lower loop orders. A security factor~10 has been applied
        to the uncertainties.}
\end{sidewaystable}



\section{\label{app::Z2OS}$\ZOS$ to three loops}

In this appendix we provide results for the coefficients of
$\ZOS$ as defined in Eq.~(\ref{eq::deltaZ2}) up to three loops
including higher order terms in $\epsilon$: the $n$-loop
expression contains terms up to order $\epsilon^{4-n}$.
Note that in Eq.~(\ref{eq::deltaZ2}) the quark mass $M$ is renormalized
on-shell but $\alpha_s$ is bare. Our results read
\begin{eqnarray}
  \delta Z_2^{(1)} &=&
 \big(\frac{\zeta _3}{3}-\frac{3 \pi ^4}{640}-\frac{\pi
   ^2}{6}-8\big) \epsilon ^3 C_F+\big(\frac{\zeta _3}{4}-\frac{\pi
   ^2}{12}-4\big) \epsilon ^2 C_F+\big(-2-\frac{\pi ^2}{16}\big)
   \epsilon  C_F  
\nonumber\\&& 
-\frac{3 C_F}{4 \epsilon }-C_F
\,,
\end{eqnarray}
\begin{eqnarray}
  \delta Z_2^{(2)} &=&
 \epsilon  \big(C_A C_F \big(12 a_4+\frac{199 \zeta
   _3}{24}-\frac{7 \pi ^4}{40}+\frac{227 \pi
   ^2}{384}-\frac{4241}{256}+\frac{\log ^4(2)}{2}
\nonumber\\&& 
+\pi ^2 \log
   ^2(2)
-\frac{23}{8} \pi ^2 \log (2)\big)
\nonumber\\&& 
+C_F^2 \big(-24 a_4-\frac{297
   \zeta _3}{16}+\frac{7 \pi ^4}{20}-\frac{339 \pi
   ^2}{128}+\frac{211}{256}-\log ^4(2)
\nonumber\\&& 
-2 \pi ^2 \log ^2(2)+\frac{23}{4} \pi
   ^2 \log (2)\big)
\nonumber\\&& 
+C_F \big(-\frac{43 \zeta _3}{6}-\frac{437 \pi
   ^2}{288}+\frac{20275}{1728}
\nonumber\\&& 
+2 \pi ^2 \log (2)\big) T_F
   n_h+\big(\frac{11 \zeta _3}{12}+\frac{15 \pi
   ^2}{32}+\frac{369}{64}\big) C_F T_F n_l\big)
\nonumber\\&& 
+\epsilon ^2 \big(C_A
   C_F \big(69 a_4+72 a_5
-\frac{11 \pi ^2 \zeta _3}{8}+\frac{2561 \zeta
   _3}{96}-\frac{609 \zeta _5}{8}-\frac{7229 \pi ^4}{11520}
\nonumber\\&& 
+\frac{2005 \pi
   ^2}{768}
-\frac{30163}{512}-\frac{3 \log ^5(2)}{5}+\frac{23 \log
   ^4(2)}{8}-2 \pi ^2 \log ^3(2)+\frac{23}{4} \pi ^2 \log
   ^2(2)
\nonumber\\&& 
+\frac{13}{30} \pi ^4 \log (2)
-\frac{41}{4} \pi ^2 \log
   (2)\big)
\nonumber\\&& 
+C_F^2 \big(-138 a_4-144 a_5+\frac{11 \pi ^2 \zeta
   _3}{4}-\frac{2069 \zeta _3}{32}+\frac{609 \zeta _5}{4}+\frac{3901 \pi
   ^4}{3840}
\nonumber\\&& 
-\frac{8851 \pi ^2}{768}+\frac{4889}{512}+\frac{6 \log
   ^5(2)}{5}-\frac{23 \log ^4(2)}{4}+4 \pi ^2 \log ^3(2)-\frac{23}{2} \pi ^2
   \log ^2(2)
\nonumber\\&& 
-\frac{13}{15} \pi ^4 \log (2)+\frac{41}{2} \pi ^2 \log
   (2)\big)
\nonumber\\&& 
+C_F T_F n_h \big(-48 a_4-\frac{2413 \zeta _3}{72}+\frac{47
   \pi ^4}{160}-\frac{8509 \pi ^2}{1728}
\nonumber\\&& 
+\frac{450395}{10368}-2 \log ^4(2)-4
   \pi ^2 \log ^2(2)+\frac{19}{2} \pi ^2 \log (2)\big)
\nonumber\\&& 
+\big(\frac{33
   \zeta _3}{8}+\frac{101 \pi ^4}{960}+\frac{295 \pi
   ^2}{192}+\frac{2259}{128}\big) C_F T_F n_l\big)
\nonumber\\&& 
+C_A C_F \big(\frac{3
   \zeta _3}{4}+\frac{49 \pi ^2}{192}-\frac{803}{128}-\frac{1}{2} \pi ^2
   \log (2)\big)
\nonumber\\&& 
+C_F^2 \big(-\frac{3
   \zeta _3}{2}-\frac{49 \pi ^2}{64}+\frac{433}{128}+\pi ^2 \log
   (2)\big)+\big(\frac{1139}{288}-\frac{7 \pi ^2}{24}\big) C_F T_F
   n_h
\nonumber\\&& 
+\big(\frac{59}{32}+\frac{5 \pi ^2}{48}\big) C_F T_F n_l
\nonumber\\&& 
+\frac{-\frac{11 C_A C_F}{32}+\frac{1}{4} C_F T_F
   n_h+\frac{1}{8} C_F T_F n_l+\frac{9 C_F^2}{32}}{\epsilon
   ^2}
\nonumber\\&& 
+\frac{-\frac{101 C_A C_F}{64}+\frac{19}{48} C_F T_F n_h+\frac{9}{16}
   C_F T_F n_l+\frac{51 C_F^2}{64}}{\epsilon }
\,,
\end{eqnarray}
\begin{eqnarray}
  \delta Z_2^{(3)} &=&
\big(-10 a_4+\frac{\pi ^2 \zeta _3}{8}-\frac{739
   \zeta _3}{128}-\frac{5 \zeta _5}{16}-\frac{5 \log ^4(2)}{12}+3 \pi ^2
   \log ^2(2)
\nonumber\\&& 
+\frac{685}{48} \pi ^2 \log (2)-\frac{41 \pi
   ^4}{120}-\frac{58321 \pi ^2}{9216}-\frac{10823}{3072}\big) C_F^3
\nonumber\\&& 
+C_A
   \big(-\frac{319 a_4}{6}-\frac{45 \pi ^2 \zeta _3}{16}-\frac{19981 \zeta
   _3}{384}+\frac{145 \zeta _5}{16}-\frac{319 \log
   ^4(2)}{144}-\frac{499}{72} \pi ^2 \log ^2(2)
\nonumber\\&& 
+\frac{2281}{288} \pi ^2 \log
   (2)+\frac{20053 \pi ^4}{17280}-\frac{15053 \pi
   ^2}{9216}+\frac{150871}{9216}\big) C_F^2
\nonumber\\&& 
+T_F n_l \big(\frac{64
   a_4}{3}+\frac{1661 \zeta _3}{96}+\frac{8 \log ^4(2)}{9}+\frac{16}{9} \pi
   ^2 \log ^2(2)-\frac{58}{9} \pi ^2 \log (2)-\frac{733 \pi
   ^4}{2160}
\nonumber\\&& 
+\frac{6931 \pi ^2}{2304}-\frac{3773}{2304}\big) C_F^2
\nonumber\\&& 
+T_F
   n_h \big(28 a_4+\frac{5327 \zeta _3}{288}+\frac{7 \log
   ^4(2)}{6}+\frac{5}{6} \pi ^2 \log ^2(2)-\frac{31}{9} \pi ^2 \log
   (2)
\nonumber\\&& 
-\frac{137 \pi ^4}{720}+\frac{25223 \pi
   ^2}{20736}-\frac{78967}{6912}\big) C_F^2
\nonumber\\&& 
+T_F^2 n_l^2
   \big(-\frac{37 \zeta _3}{36}-\frac{23 \pi
   ^2}{48}-\frac{4025}{972}\big) C_F
\nonumber\\&& 
+T_F^2 n_h n_l \big(\frac{49 \zeta
   _3}{12}-\frac{4}{3} \pi ^2 \log (2)+\frac{77 \pi
   ^2}{72}-\frac{1168}{81}\big) C_F
\nonumber\\&& 
+T_F^2 n_h^2 \big(\frac{85 \zeta
   _3}{12}-\frac{4}{3} \pi ^2 \log (2)+\frac{767 \pi
   ^2}{720}-\frac{6887}{648}\big) C_F
\nonumber\\&& 
+C_A T_F n_h \big(-16 a_4+\xi 
   \big(-\frac{7 \zeta _3}{192}+\frac{\pi
   ^2}{256}+\frac{407}{1728}\big)+\frac{11 \pi ^2 \zeta _3}{48}-\frac{3359
   \zeta _3}{144}
-\frac{15 \zeta _5}{16}
\nonumber\\&& 
-\frac{2 \log ^4(2)}{3}-\frac{1}{3}
   \pi ^2 \log ^2(2)+\frac{521}{36} \pi ^2 \log (2)+\frac{5 \pi
   ^4}{72}-\frac{105359 \pi ^2}{10368}+\frac{32257}{648}\big) C_F
\nonumber\\&& 
+C_A T_F
   n_l \big(-\frac{32 a_4}{3}-\frac{301 \zeta _3}{72}-\frac{4 \log
   ^4(2)}{9}-\frac{8}{9} \pi ^2 \log ^2(2)+\frac{29}{9} \pi ^2 \log
   (2)
\nonumber\\&& 
+\frac{29 \pi ^4}{216}
+\frac{2413 \pi
   ^2}{3456}+\frac{416405}{15552}\big) C_F
\nonumber\\&& 
+C_A^2 \big(\frac{349
   a_4}{12}+\frac{127 \pi ^2 \zeta _3}{72}+\frac{3623 \zeta _3}{144}+\xi 
   \big(\frac{\pi ^2 \zeta _3}{144}-\frac{13 \zeta _3}{256}+\frac{7 \zeta
   _5}{384}+\frac{17 \pi ^4}{27648}
\nonumber\\&& 
-\frac{\pi
   ^2}{256}-\frac{13}{768}\big)-\frac{37 \zeta _5}{6}+\frac{349 \log
   ^4(2)}{288}+\frac{391}{144} \pi ^2 \log ^2(2)-\frac{271}{36} \pi ^2 \log
   (2)
\nonumber\\&& 
-\frac{10811 \pi ^4}{23040}-\frac{107 \pi
   ^2}{864}-\frac{2551697}{62208}\big) C_F
\nonumber\\&& 
+\frac{1}{\epsilon
   ^3}\big[-\frac{9
   C_F^3}{128}+\frac{33}{128} C_A C_F^2-\frac{3}{16} T_F n_h
   C_F^2-\frac{3}{32} T_F n_l C_F^2-\frac{121}{576} C_A^2
   C_F
\nonumber\\&& 
-\frac{1}{12} T_F^2 n_h^2 C_F-\frac{1}{36} T_F^2 n_l^2
   C_F+\frac{11}{72} C_A T_F n_l C_F-\frac{1}{12} T_F^2 n_h n_l C_F
\nonumber\\&& 
+C_A
   T_F n_h \big(\frac{15}{64}-\frac{\xi }{192}\big) C_F\big]
\nonumber\\&& 
+\frac{1}{\epsilon ^2}\big[-\frac{81 C_F^3}{256}+\frac{1217}{768} C_A
   C_F^2-\frac{91}{192} T_F n_h C_F^2-\frac{103}{192} T_F n_l
   C_F^2-\frac{1501}{864} C_A^2 C_F
\nonumber\\&& 
-\frac{5}{36} T_F^2 n_h^2
   C_F-\frac{23}{108} T_F^2 n_l^2 C_F+\frac{1069}{864} C_A T_F n_l
   C_F-\frac{7}{18} T_F^2 n_h n_l C_F
\nonumber\\&& 
+C_A T_F n_h \big(\frac{\xi
   }{64}+\frac{353}{288}\big) C_F\big]
\nonumber\\&& 
+\frac{1}{\epsilon }\big[\big(\frac{9 \zeta
   _3}{8}-\frac{3}{4} \pi ^2 \log (2)+\frac{303 \pi
   ^2}{512}-\frac{1039}{512}\big) C_F^3+T_F n_l \big(\frac{3 \zeta
   _3}{4}-\frac{2}{3} \pi ^2 \log (2)
\nonumber\\&& 
+\frac{175 \pi
   ^2}{384}-\frac{351}{128}\big) C_F^2+T_F n_h \big(\zeta
   _3-\frac{2}{3} \pi ^2 \log (2)+\frac{143 \pi
   ^2}{192}-\frac{1525}{384}\big) C_F^2
\nonumber\\&& 
+C_A \big(-\frac{27 \zeta
   _3}{8}+\frac{53}{24} \pi ^2 \log (2)-\frac{2549 \pi
   ^2}{1536}+\frac{14887}{1536}\big) C_F^2
\nonumber\\&& 
+C_A^2 \big(\xi 
   \big(-\frac{3 \zeta _3}{256}+\frac{\pi
   ^4}{4320}-\frac{1}{768}\big)
\nonumber\\&& 
+\frac{173 \zeta _3}{128}-\frac{11}{12} \pi
   ^2 \log (2)-\frac{\pi ^4}{1080}+\frac{1199 \pi
   ^2}{2304}-\frac{55945}{5184}\big) C_F
\nonumber\\&& 
+C_A T_F n_h
   \big(\big(-\frac{35}{576}-\frac{\pi ^2}{768}\big) \xi -\frac{\zeta
   _3}{2}+\frac{1}{3} \pi ^2 \log (2)-\frac{1753 \pi
   ^2}{2304}+\frac{503}{48}\big) C_F
\nonumber\\&& 
+C_A T_F n_l \big(-\frac{\zeta
   _3}{4}+\frac{1}{3} \pi ^2 \log (2)-\frac{5 \pi
   ^2}{288}+\frac{550}{81}\big) C_F
+T_F^2 n_h^2
   \big(-\frac{131}{54}+\frac{29 \pi ^2}{144}\big) C_F
\nonumber\\&& 
+T_F^2 n_h n_l
   \big(-\frac{31}{9}+\frac{7 \pi ^2}{48}\big) C_F+T_F^2 n_l^2
   \big(-\frac{325}{324}-\frac{\pi ^2}{16}\big) C_F\big]
\nonumber\\&& 
+\epsilon 
   \big(\big(\frac{29 \zeta _3^2}{32}+14 \pi ^2 \log (2) \zeta
   _3+\frac{5267 \pi ^2 \zeta _3}{288}-\frac{69511 \zeta _3}{256}-\frac{4267
   a_4}{6}-\frac{116 a_5}{3}
\nonumber\\&& 
-\frac{1403 \zeta _5}{16}+\frac{29 \log
   ^5(2)}{90}+\frac{2}{3} \pi ^2 \log ^4(2)-\frac{4267 \log
   ^4(2)}{144}-\frac{196}{27} \pi ^2 \log ^3(2)
\nonumber\\&& 
-\frac{2}{3} \pi ^4 \log
   ^2(2)-\frac{3997}{72} \pi ^2 \log ^2(2)-\frac{2351 \pi ^4 \log
   (2)}{2160}+\frac{1345}{8} \pi ^2 \log (2)-\frac{899 \pi
   ^6}{5670}
\nonumber\\&& 
+\frac{450893 \pi ^4}{552960}+16 a_4 \pi ^2-\frac{56455 \pi
   ^2}{2048}-\frac{108677}{6144}\big) C_F^3
\nonumber\\&& 
+T_F n_l \big(\frac{1856
   a_4}{9}+\frac{512 a_5}{3}-\frac{69 \pi ^2 \zeta _3}{16}+\frac{57581 \zeta
   _3}{576}-\frac{2245 \zeta _5}{12}-\frac{64 \log ^5(2)}{45}
\nonumber\\&& 
+\frac{232 \log
   ^4(2)}{27}-\frac{128}{27} \pi ^2 \log ^3(2)+\frac{464}{27} \pi ^2 \log
   ^2(2)+\frac{139}{270} \pi ^4 \log (2)-\frac{908}{27} \pi ^2 \log
   (2)
\nonumber\\&& 
-\frac{610451 \pi ^4}{414720}+\frac{251107 \pi
   ^2}{13824}-\frac{36677}{13824}\big) C_F^2
\nonumber\\&& 
+T_F n_h \big(\frac{539
   a_4}{6}+168 a_5-\frac{287 \pi ^2 \zeta _3}{24}+\frac{1087 \zeta
   _3}{96}-\frac{899 \zeta _5}{8}-\frac{7 \log ^5(2)}{5}
\nonumber\\&& 
+\frac{539 \log
   ^4(2)}{144}-\frac{5}{3} \pi ^2 \log ^3(2)-\frac{683}{144} \pi ^2 \log
   ^2(2)+\frac{247}{180} \pi ^4 \log (2)+\frac{673}{24} \pi ^2 \log
   (2)
\nonumber\\&& 
-\frac{32857 \pi ^4}{69120}-\frac{341735 \pi
   ^2}{41472}-\frac{143029}{13824}\big) C_F^2
\nonumber\\&& 
+C_A \big(-\frac{143 \zeta
   _3^2}{4}+\frac{21}{4} \pi ^2 \log (2) \zeta _3-\frac{5777 \pi ^2 \zeta
   _3}{384}-\frac{188083 \zeta _3}{2304}-\frac{3787 a_4}{36}
\nonumber\\&& 
-\frac{1441
   a_5}{3}+\frac{254581 \zeta _5}{384}+\frac{1441 \log
   ^5(2)}{360}+\frac{1}{4} \pi ^2 \log ^4(2)-\frac{3787 \log
   ^4(2)}{864}
\nonumber\\&& 
+\frac{2089}{108} \pi ^2 \log ^3(2)-\frac{1}{4} \pi ^4 \log
   ^2(2)-\frac{25441}{432} \pi ^2 \log ^2(2)-\frac{559}{540} \pi ^4 \log
   (2)
\nonumber\\&& 
-\frac{5623}{216} \pi ^2 \log (2)-\frac{27331 \pi
   ^6}{181440}+\frac{11810161 \pi ^4}{1658880}+6 a_4 \pi ^2
\nonumber\\&& 
-\frac{1265393
   \pi ^2}{55296}+\frac{4824655}{55296}\big) C_F^2
\nonumber\\&& 
+T_F^2 n_l^2
   \big(-\frac{851 \zeta _3}{108}-\frac{535 \pi ^4}{3456}-\frac{325 \pi
   ^2}{144}-\frac{23789}{1458}\big) C_F
\nonumber\\&& 
+T_F^2 n_h n_l \big(\frac{128
   a_4}{3}+\frac{595 \zeta _3}{18}+\frac{16 \log ^4(2)}{9}+\frac{32}{9} \pi
   ^2 \log ^2(2)-\frac{92}{9} \pi ^2 \log (2)
\nonumber\\&& 
-\frac{341 \pi
   ^4}{3456}+\frac{145 \pi ^2}{36}-\frac{17740}{243}\big) C_F
\nonumber\\&& 
+T_F^2
   n_h^2 \big(\frac{152 a_4}{3}+\frac{25723 \zeta _3}{720}+\frac{19 \log
   ^4(2)}{9}+\frac{17}{9} \pi ^2 \log ^2(2)-\frac{33}{5} \pi ^2 \log
   (2)
\nonumber\\&& 
-\frac{391 \pi ^4}{1152}+\frac{83227 \pi
   ^2}{16200}-\frac{1023397}{19440}\big) C_F
\nonumber\\&& 
+C_A^2 \big(\frac{7451
   \zeta _3^2}{384}-\frac{49}{8} \pi ^2 \log (2) \zeta _3+\frac{3299 \pi
   ^2 \zeta _3}{4608}+\frac{236171 \zeta _3}{3456}+\frac{4147
   a_4}{18}+\frac{1499 a_5}{6}
\nonumber\\&& 
+\xi  \big(\frac{25 \zeta
   _3^2}{384}+\frac{77 \pi ^2 \zeta _3}{9216}-\frac{63 \zeta
   _3}{256}-\frac{149 \zeta _5}{768}+\frac{49 \pi ^6}{51840}+\frac{383 \pi
   ^4}{138240}-\frac{35 \pi ^2}{1024}-\frac{35}{256}\big)
\nonumber\\&& 
-\frac{77297
   \zeta _5}{256}-\frac{1499 \log ^5(2)}{720}-\frac{7}{24} \pi ^2 \log
   ^4(2)+\frac{4147 \log ^4(2)}{432}-\frac{1697}{216} \pi ^2 \log
   ^3(2)
\nonumber\\&& 
+\frac{7}{24} \pi ^4 \log ^2(2)+\frac{4679}{108} \pi ^2 \log
   ^2(2)+\frac{6823 \pi ^4 \log (2)}{8640}-\frac{25069}{864} \pi ^2 \log
   (2)
\nonumber\\&& 
+\frac{45047 \pi ^6}{362880}-\frac{2973217 \pi ^4}{829440}-7 a_4 \pi
   ^2+\frac{26425 \pi ^2}{20736}-\frac{74570603}{373248}\big) C_F
\nonumber\\&& 
+C_A T_F
   n_l \big(-\frac{928 a_4}{9}-\frac{256 a_5}{3}+\frac{77 \pi ^2 \zeta
   _3}{48}-\frac{11159 \zeta _3}{864}+\frac{2077 \zeta _5}{24}+\frac{32 \log
   ^5(2)}{45}
\nonumber\\&& 
-\frac{116 \log ^4(2)}{27}+\frac{64}{27} \pi ^2 \log
   ^3(2)-\frac{232}{27} \pi ^2 \log ^2(2)-\frac{139}{540} \pi ^4 \log
   (2)+\frac{454}{27} \pi ^2 \log (2)
\nonumber\\&& 
+\frac{122131 \pi
   ^4}{103680}+\frac{10535 \pi ^2}{2592}+\frac{10805903}{93312}\big)
   C_F
\nonumber\\&& 
+C_A T_F n_h \big(\frac{181 \zeta _3^2}{32}+7 \pi ^2 \log (2) \zeta
   _3+\frac{1823 \pi ^2 \zeta _3}{192}-\frac{357881 \zeta
   _3}{1152}-\frac{5855 a_4}{12}-96 a_5
\nonumber\\&& 
+\xi  \big(\frac{7 \zeta
   _3}{64}+\frac{173 \pi ^4}{92160}-\frac{35 \pi
   ^2}{2304}-\frac{4859}{5184}\big)+\frac{2697 \zeta _5}{64}+\frac{4 \log
   ^5(2)}{5}+\frac{1}{3} \pi ^2 \log ^4(2)
\nonumber\\&& 
-\frac{5855 \log
   ^4(2)}{288}+\frac{2}{3} \pi ^2 \log ^3(2)-\frac{1}{3} \pi ^4 \log
   ^2(2)-\frac{21349}{288} \pi ^2 \log ^2(2)-\frac{149}{180} \pi ^4 \log
   (2)
\nonumber\\&& 
+\frac{3973}{36} \pi ^2 \log (2)-\frac{1501 \pi
   ^6}{15120}+\frac{173713 \pi ^4}{55296}+8 a_4 \pi ^2
\nonumber\\&& 
-\frac{163981 \pi
   ^2}{5184}+\frac{1004165}{3888}\big) C_F\big)
\,,
\label{eq::deltaZ2OS_3}
\end{eqnarray}
where $C_F=(N_c^2-1)/(2N_c)$ and $C_A=N_c$
are the eigenvalues of the quadratic Casimir operators
of the fundamental and adjoint representation for the SU$(N_c)$ colour group,
respectively, $T=1/2$ is the index of the fundamental representation, and
$n_l$ and $n_h$ count the number of massless and massive (with mass $M$)
quarks. It is convenient to keep the variable $n_h$ as a parameter although in
our case we have $n_h=1$.  Computer-readable expressions of $\delta
Z_2^{(1)}$, $\delta Z_2^{(2)}$ and $\delta Z_2^{(3)}$ can be found
in~\cite{progdata}.


\end{appendix}

\end{document}